\documentclass[a4paper]{llncs}
\pdfoutput=1
\usepackage{graphics}
\usepackage{listings}
\usepackage[colorinlistoftodos]{todonotes}
\usepackage[misc]{ifsym}
\begin{document}

\title{Domain Specific Semantic Validation of Schema.org Annotations}
%
%
\author{Umutcan \c{S}im\c{s}ek\Letter \and Elias K\"{a}rle \and
Omar Holzknecht \and Dieter Fensel}
%
%
%
\institute{STI Innsbruck, University of Innsbruck, Technikerstrasse 21a 6020 Innsbruck, Austria\\
\email{\{umutcan.simsek, elias.kaerle, omar.holzknecht, dieter.fensel\}@sti2.at},\\
\texttt{http://www.sti2.at}}
\maketitle              

\begin{abstract}
Since its unveiling in 2011, schema.org has become the de facto standard for publishing semantically described structured data on the web, typically in the form of web page annotations. The increasing adoption of schema.org facilitates the growth of the web of data, as well as the development of automated agents that operate on this data. Schema.org is a large heterogeneous vocabulary that covers many domains. This is obviously not a bug, but a feature, since schema.org aims to describe almost everything on the web, and the web is huge. However, the heterogeneity of schema.org may cause a side effect, which is the challenge of picking the right classes and properties for an annotation in a certain domain, as well as keeping the annotation semantically consistent. In this work, we introduce our rule based approach and an implementation of it for validating schema.org annotations from two aspects: (a) the completeness of the annotations in terms of a specified domain, (b) the semantic consistency of the values based on pre-defined rules. We demonstrate our approach in the tourism domain.
\end{abstract}

\section{Introduction}
\label{sec:introduction}

To publish structured data on the web there are a lot of collections of vocabularies and ontologies that all serve a different or overlapping purpose and appear, grow and vanish in an unpredictable manner. However, there is one initiative to provide structured data on the web which stands out by means of community adoption and distribution and became a de facto standard, which is schema.org\footnote{https://schema.org}. Schema.org was developed in 2011 by Google, Microsoft, Yahoo! and Yandex and has been supported since by a broad community and found application on millions of websites\cite{Guha:2016:SES:2886013.2844544}. Schema.org can be included into the website's source code with common technologies like Microdata, RDFa or JSON-LD. The vocabulary covers local businesses, products, events, recipes, people and much more and is adapted and supported by the big search engine providers. This naturally makes the vocabulary quite heterogeneous. The vocabulary is also semantically imperfect \cite{patel2014analyzing}. For instance classes may inherit properties improperly (e.g. a waterfall can have a telephone number) and not formally strict, but this is rather a design decision to facilitate rapid and decentralized evolution of the vocabulary. The side effect of this feature is that picking the right classes and properties for a domain can be quite challenging and low quality annotations in terms of conforming to the rules of a field (e.g. tourism) may occur. 

The World Wide Web was originally designed as an internet-based hypertext system. It contains blocks of information, the websites, which are connected via hyperlinks to other blocks of information. Due to that simple design and the open-to-all approach it rapidly evolved to be the biggest information network that ever existed. The headless web\footnote{https://paul.kinlan.me/the-headless-web/} is a layer which grows on top of the Web we know. Within this layer goods are not sold by individual producers or small retail websites, but by a few large retail platforms like Alibaba or Amazon. Rooms are not sold by hotels or destination marketing organizations (DMOs) but by a hand full of huge online travel agencies (OTAs) like booking.com or Expedia. In a not too distant future information will no longer be found on individual websites, but gathered by the search engines and presented to the searching user directly on the search engine website. So the web is, in the true sense of the word, losing its head: its graphical representation.
The data will be extracted from websites and presented to the user not only by the search engines but also by personal assistant software like Cortana, Siri, or Google Now. 
With this new layer we can observe a trend towards going-out-of-use of graphical representation and the rising necessity of structured, high quality, data. 
The data for services like Cortana or Siri is going to be collected and gathered by crawlers and only structured, machine read- and understandable data will be part of the game at that point. 
In the headless web there will be no room for unstructured but beautifully designed content. The challenge for small and medium enterprises (SME) is to bring their data into this new layer by precise, correct and complete semantic annotations on their websites. Schema.org is the vocabulary of choice to do that and hence SMEs need a way to produce schema.org annotations in a correct way and a tool to validate those annotations.

This paper describes such a method to define domain specific subsets of the schema.org vocabulary with enriched semantics and also introduces the tool we provide in order to validate the semantics of domain specific structured data annotated with schema.org on websites. Depending on the domain, a subset of schema.org classes and properties will be selected and a set of rules will be defined by a domain expert - which is the foundation of the validation process. From there on users can validate their own annotations and websites based on the domain specific subset and the validation rules defined by the domain expert.

The remainder of this paper is organized as follows:  Section \ref{sec:related-work} compares the described approach with related work. Section \ref{sec:technical-approach} describes our method which includes a domain definition and validation approach and a tool that implements it. Section \ref{sec:use-case} shows the approach in action and section \ref{sec:conclusion} gives an outlook to future work and concludes the paper.

\section{Related Work and Motivation}
\label{sec:related-work}
While the adoption of schema.org has been increasing\cite{Meusel:2015:WSA:2797115.2797124}, the conformance of the schema.org annotations to the vocabulary specification is still questionable. A large scale study on the usage of schema.org in the tourism domain \cite{karle2016there} shows that the schema.org vocabulary is mostly used incorrectly or missing fundamental properties (e.g. many hotels do not have address information in their annotations). The issue of completeness for the schema.org annotations occurs due to the size of the vocabulary and the lack of guidance for adopters to decide which classes and properties to use. In addition to this issue, there is also the semantic consistency issue (e.g. consistency between the country and the country code of a phone number) for annotations that is not possible to capture with the prominent validation tools like the Google Structured Data Testing Tool\footnote{https://search.google.com/
	structured-data/testing-tool}.

Given the developments about the new layer on top of the web, providing well formed and semantically consistent structured data on the web is more important than ever. 
Therefore, we propose an approach, that allows us to obtain a specific subset of the schema.org vocabulary containing important classes and properties for a domain and to validate the annotations based on pre-defined rules to ensure the completeness and the semantic correctness of the data. 

The related work to our approach comes mostly from the RDF validation domain. An approach described in \cite{Fürber2010} applies SPIN Rules for domain independent detection of certain data quality problems namely, inconsistency (i.e. inconsistent representation of the data, functional dependency and referential integrity), comprehensibility (i.e. ambiguity of the data), heterogeneity and redundancy. An approach \cite{Simister2013} presented in the RDF Validation Workshop \cite{rdfval2013} proposes a simple mechanism for declaring the properties to be used for a class and a SPARQL based extension for defining more complex constraints. Parallel to the RDF Validation Workshop results, there have been an increased development of new RDF validation methods. Shape Expressions (ShEx) \cite{Prudhommeaux2014} is a domain specific language for validating and transforming RDF Data. Similar to ShEx, RDF Data Shapes Working Group has been developing the Shapes Constraint Language (SHACL) \cite{Knublauch2016} for describing and validating RDF graphs. SHACL allows us to define constraints targeting specific nodes in a data graph based on their type, identifier, or a filtering SPARQL query. It is currently investigated that at what level SHEx can be represented in SHACL, based on the identified similarities and differences \footnote{http://shex.io/primer/\#rel-to-shacl}.
The rule-based validation of RDF data is an emerging field, mostly focused around the re-use of prominent standards like SPARQL. All of the aforementioned validation approaches are somewhat compatible with SPARQL. Our approach shows similarities with aforementioned approaches in terms of using rules for checking consistency of the data and defining constraints over classes. The works in \cite{Prudhommeaux2014} and \cite{Knublauch2016}  allow us to define "shapes" that constraint types and instances in terms of subset of properties and expected types for those properties as well as nested shapes. 

We introduce the notion of "domain" and a simple specification of it for schema.org, which adopts a similar nested definition of constraints that restricts classes and properties in relation to other classes of which they are expected types. The concept of selecting a subset of schema.org appears in \cite{Khalili2013}, but to the best of our knowledge, the domain selection of the editor described there is limited to the selection of classes. We propose a different domain specification approach including selecting a subset of properties and restricting the range of those properties to a subset of subclasses of the range defined by schema.org. The importance of this restriction is described in Section \ref{sec:dom-def} in more detail. Additionally, our validator brings domain definition and semantic consistency rules together in one holistic tool.

In order to show a concrete example of our motivation for domain specific validation, we can consider annotation of an event. The \textit{Event} class of schema.org vocabulary contains 38 properties including the ones inherited from the \textit{Thing} class. Even though this number seems not too high, the properties whose range is a complex type makes the annotation size unmanageable. Let us take only one property of the Event class into account: organizer. This property can have values in the \textit{Organization} class. If a user starts to annotate an event and its organizer, she will soon realize that the \textit{Organization} class itself offers 50 properties. The amount of properties and classes the user needs to deal with explodes as we continue. When we define a domain, we can select a subset of properties of the \textit{Organization} class as the value of the organizer property, for instance, to only name and url. This restriction of classes when they are the value of a certain property will give a clear idea to the user who creates schema.org annotations.

\section{Method}
\label{sec:technical-approach}
In this section, we explain our approach in detail and demonstrate the web based tool \footnote{http://sdo-validator.sti2.at} that implements it.

Our approach consists of two main parts. First, the definition of a domain by selecting a subset of classes and properties (Section \ref{sec:dom-def}) as well as a set of semantic validation rules (Section \ref{sec:rule-def}). Second, the creation and validation of  a schema.org annotation in terms of its completeness regarding the defined domain and semantic consistency based on the validation rules (Section \ref{sec:val}). 

\subsection{Domain Definition}
\label{sec:dom-def}
A domain expert, who has an extensive knowledge in a certain field (e.g. tourism), defines a domain by selecting a subset of the schema.org vocabulary, the classes and properties, which is relevant to a certain domain. Moreover, it can be specified whether a property is required for a concept or allowed to have multiple values. The domain definition process consists of the following steps: First, the domain expert selects a subset of schema.org classes. Second, she specifies the allowed properties for the selected classes, as well as whether they are optional or allowed to have multiple values. In step three, for every property added into the domain, she selects the expected types of the property. She continues the domain specification by recursively following the aforementioned steps for complex types (e.g. If the address property of a Hotel is included to the domain and its expected value type is PostalAddress, the same process should be applied also for the PostalAdress class) until the domain is complete.

\begin{figure}
	\centering
	\includegraphics[width=\textwidth]{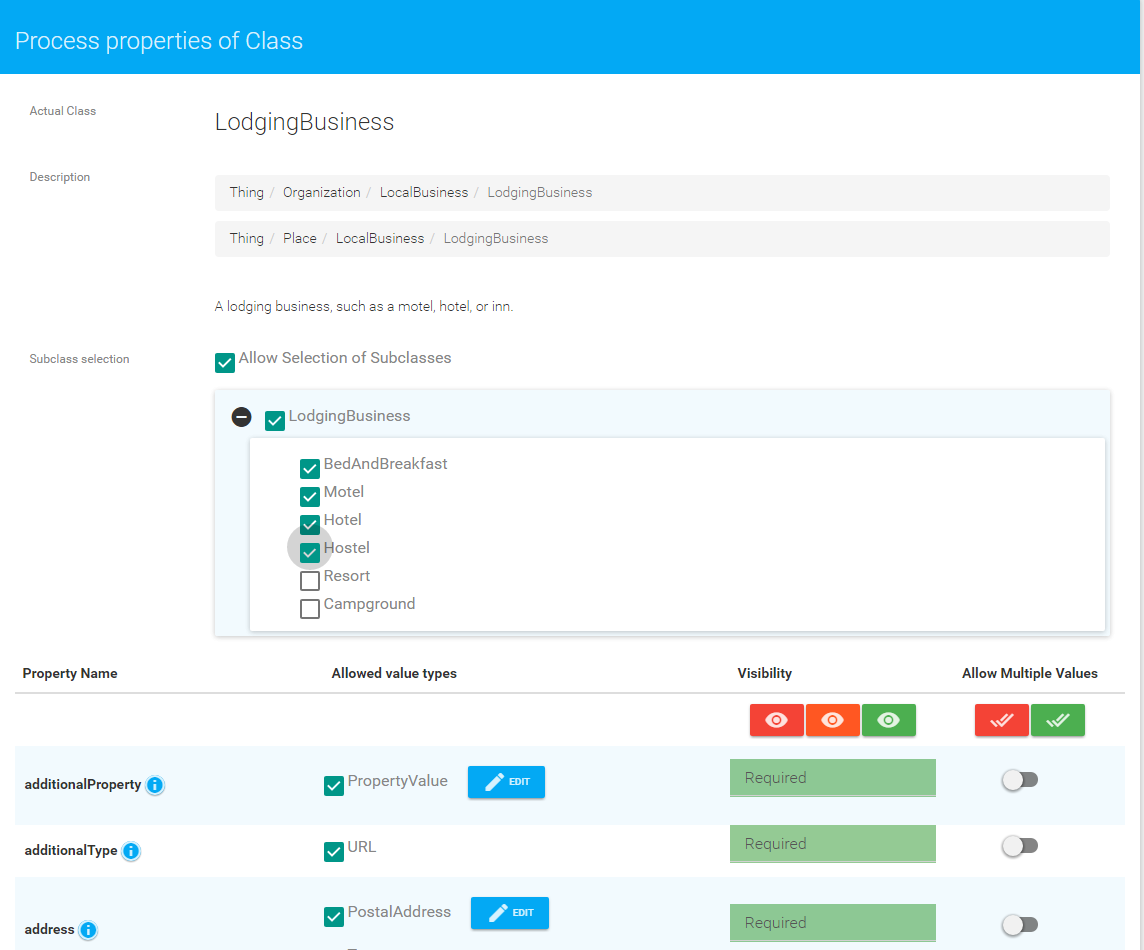}
	\caption{A screenshot from the domain definition interface. Here, a domain expert can select a subset of properties and define restrictions on them and their expected types}
	\label{fig:ddi-2}
\end{figure}

In order to facilitate the domain definition, we developed the Domain Definition Interface (Figure \ref{fig:ddi-2}) as a part of our tool. 
The aforementioned steps can be applied via the interface to create a domain. After the domain expert completes the domain, the tool generates a JSON file which contains the domain specification.

A domain specification consists of classes, that contains properties whose expected values can be in unrestricted classes (i.e. schema.org/Class) and restricted classes (e.g. a class with only a subset of its properties). Every restricted class is based on a schema.org/Class. The expected types of a property can also be restricted to a certain subset of their subclasses. Being able to restrict expected types to a subset of subclasses would be especially useful for properties like \textit{schema.org/potentialAction}, since its range is the Action class which is the most generic action. However, for a specific domain, a certain class may be required to have more specific actions as its potential action (e.g. The \textit{schema.org/potentialAction} of the\textit{ schema.org/HotelRoom} class may be restricted to \textit{schema.org/ReserveAction}).
A concrete example of a domain can be found in Section \ref{sec:use-case}.

\subsection{Rule Definition}
\label{sec:rule-def}
Rules are created by domain experts.  In order to define a rule, the domain expert first has to select a predefined domain or create a new one. Then she can create the set of rules applying to the defined domain. A semantic validation rule is a condition-action rule where an action is triggered when a condition is satisfied. Since these rules are used for validation, the condition part of a rule must state the condition that violates the domain requirement and the action part should contain the action that will be taken when the condition is satisfied (i.e. domain requirement is violated). Domain experts may use the concepts and properties that are allowed in the domain definition (section \ref{sec:dom-def}), Boolean and arithmetic operations as well as some predefined utility functions. In some cases, rules might require more complex processing of the data. To achieve this, domain experts can define their own utility function (e.g. a function that looks up for the international country calling code for a given country).
We introduce two different type of condition-action rules: local consistency and global consistency rules.
Local consistency rules compare the value of a property with a literal value (e.g. The floor size of a room must be greater than zero). An example of the local consistency rule is shown in Listing \ref{listing-local-rule}.

\begin{lstlisting}[caption="An informal representation of a local consistency validation rule", label=listing-local-rule, captionpos=b]
Condition:
HotelRoom.floorSize.QuantitativeValue.value <= 0
Action:
show("Floor size of a hotel room must be greater than zero.", Severity:Error)
\end{lstlisting}

A global consistency rule is involved with multiple properties. These properties can originate from the same class or from different classes. The following example explains the elements of a global consistency rule: A domain expert may want to create a validation rule that checks if the international country calling code of a telephone number is consistent with the country in the postal address. Such an informal validation rule may look like the Listing \ref{listing-global-rule}.
\newpage

\begin{lstlisting}[caption="An informal representation of a global consistency validation rule", label=listing-global-rule, captionpos=b]
Condition:
extractCountryCode(Place.telephone) != getCountryCodeByCountry (Place.address.PostalAddress.addressCountry)
Action:
show("The international country code of the phone number of the place is not consistent with the country of the address.", Severity:Error)
\end{lstlisting}

In the condition part, a utility function called "extractCountryCode" takes the value of the telephone property of a Place instance as parameter and returns the international country calling code. Another utility function called "getCountryCodeByCountry" takes the value of the \textit{addressCountry} property of a \textit{PostalAddress} instance of the same \textit{Place} instance and returns the international country calling code for the specified country. If the comparison shows that two values are not equal, the Action part is triggered. The predefined utility function "show" displays the reason and the severity of the violation. Rules not only define what is allowed or what is not, but also gives meaningful correction suggestions like "The phone number you specified does not match the mentioned country. Is that really correct?". These suggestions have to be defined in the rules as well.

\begin{figure}
	\centering
	\includegraphics[width=\textwidth]{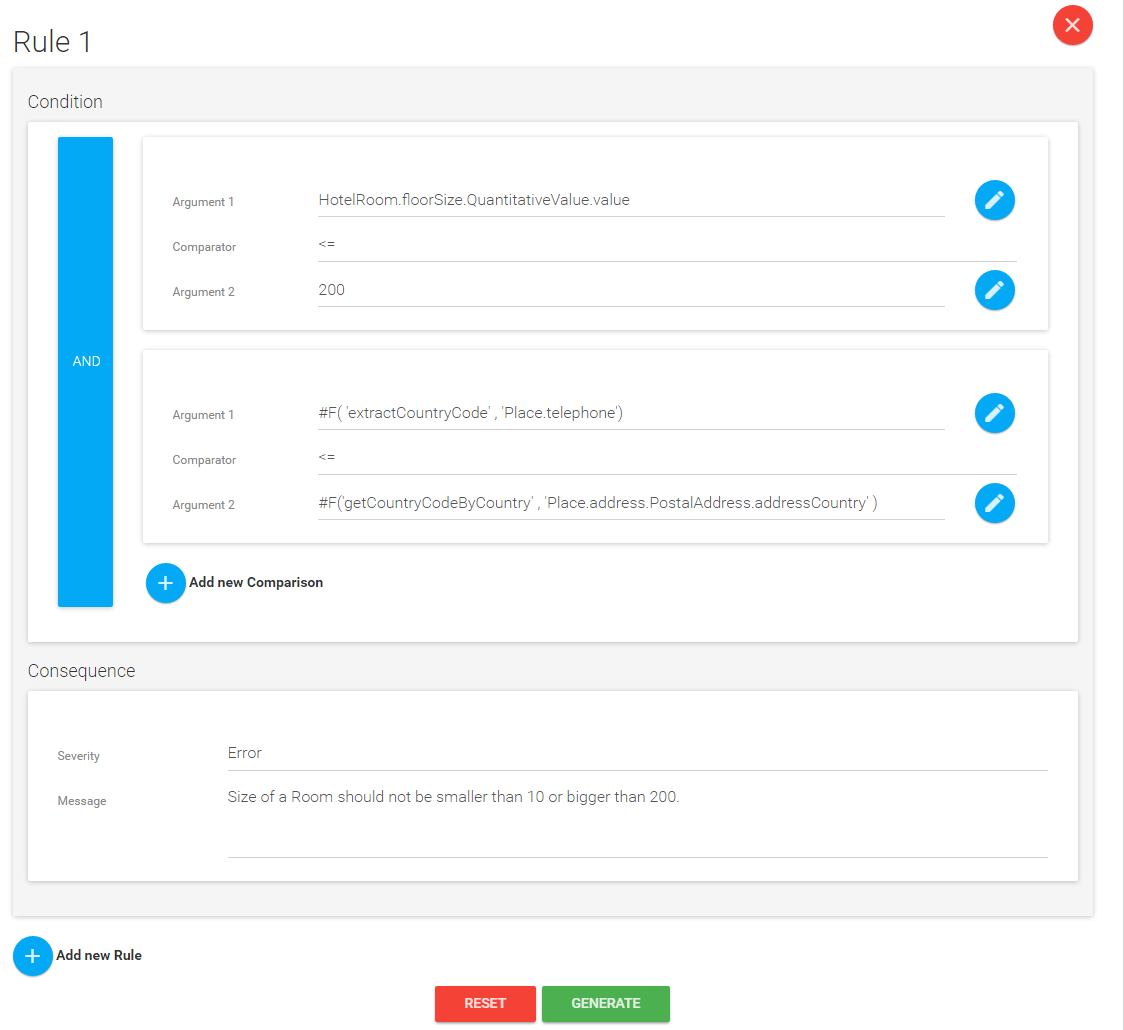}
	\caption{Prototypical interface of the Rule Designer}
	\label{fig:ruledef-1}
\end{figure}

Figure \ref{fig:ruledef-1} shows the first prototype of the rule designer, which is a form based component of our tool to enable domain experts to create semantic validation rules. Via this interface, the domain expert can create rule conditions that represent semantic inconsistencies and suitable error messages to show, in case the violation conditions are satisfied. 

\subsection{Annotation and Validation}
\label{sec:val}

In order to guide a user who wants to create an annotation in a certain domain, we generate an annotation editor based on a domain specification and ensure the completeness of the annotation. An annotation is valid in terms of completeness if it contains all required properties, none of the unspecified properties, and correct expected types for the properties defined in the domain and used in the annotation.

The annotation then can be validated for semantic consistency. The validation process iterates over all the rules defined and saves the result of the validation against each rule in a list to be presented to the user. Similar to the definition of the rules, we distinguish between local and global consistency rules. Local consistency rules consider the value of only one property, global consistency rules consider the values of several properties, check complex relations between various properties, and can go over several rules.

\begin{figure}
	\centering
	\includegraphics[width=\textwidth]{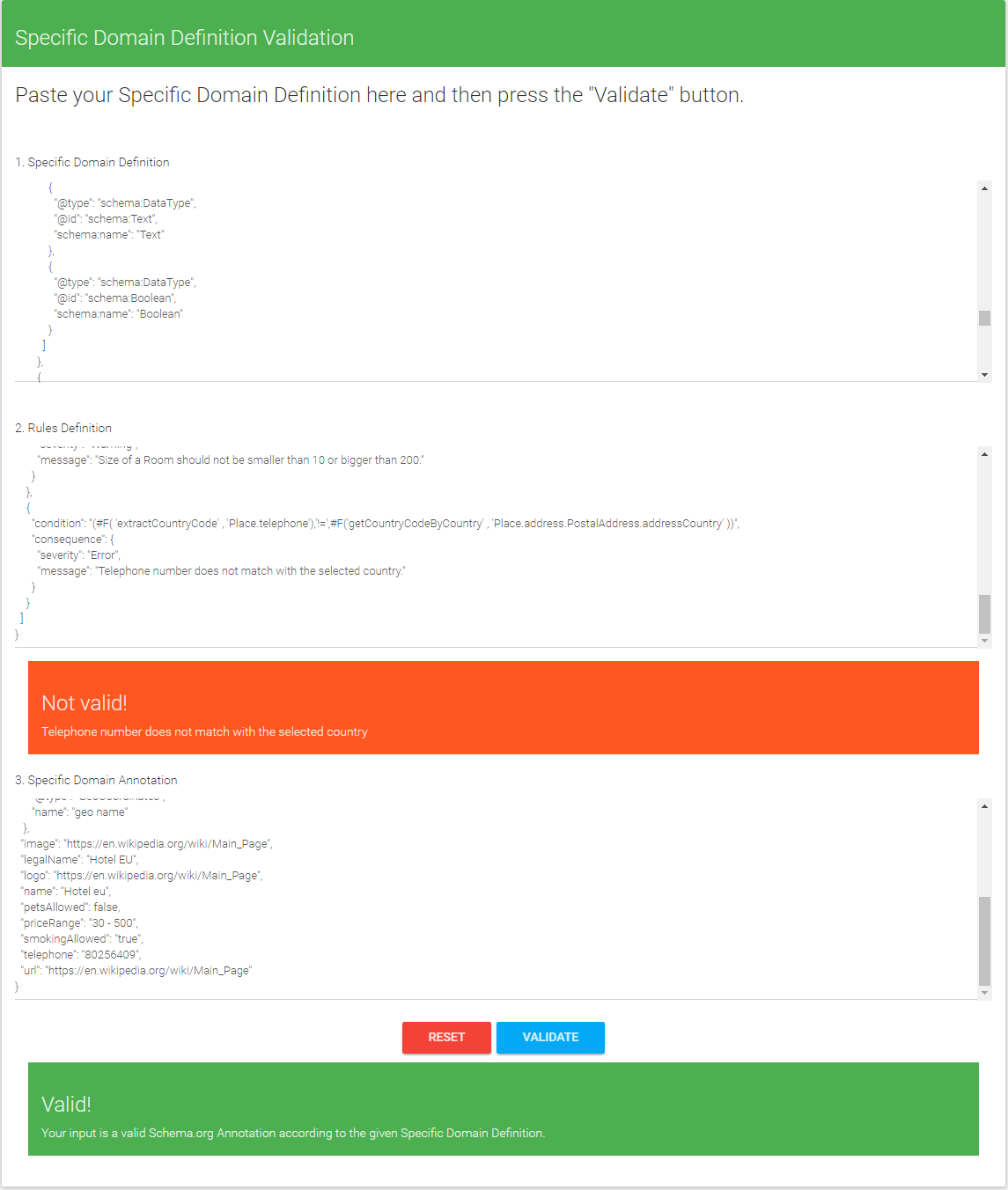}
	\caption{Validation interface}
	\label{fig:val-1}
\end{figure}

Figure \ref{fig:val-1} depicts the validation interface of our tool, which is used by the user for validation of an annotation. This interface can validate an annotation for both completeness and semantic consistency.\footnote{ For the annotations that are created via the editor based on the domain specification, only the semantic consistency validation applies.} The validator first ensures the syntactic correctness of the entries. Then it validates the completeness of the annotation. If the annotation conforms the domain specification, the validator iterates over the rules defined in the rule set and warns the user if there is any semantic inconsistency within the annotation. 

\section{Use Case: Annotation of a Lodging Business}
\label{sec:use-case}

\begin{figure}
	\centering
	\includegraphics[width=1.1\textwidth]{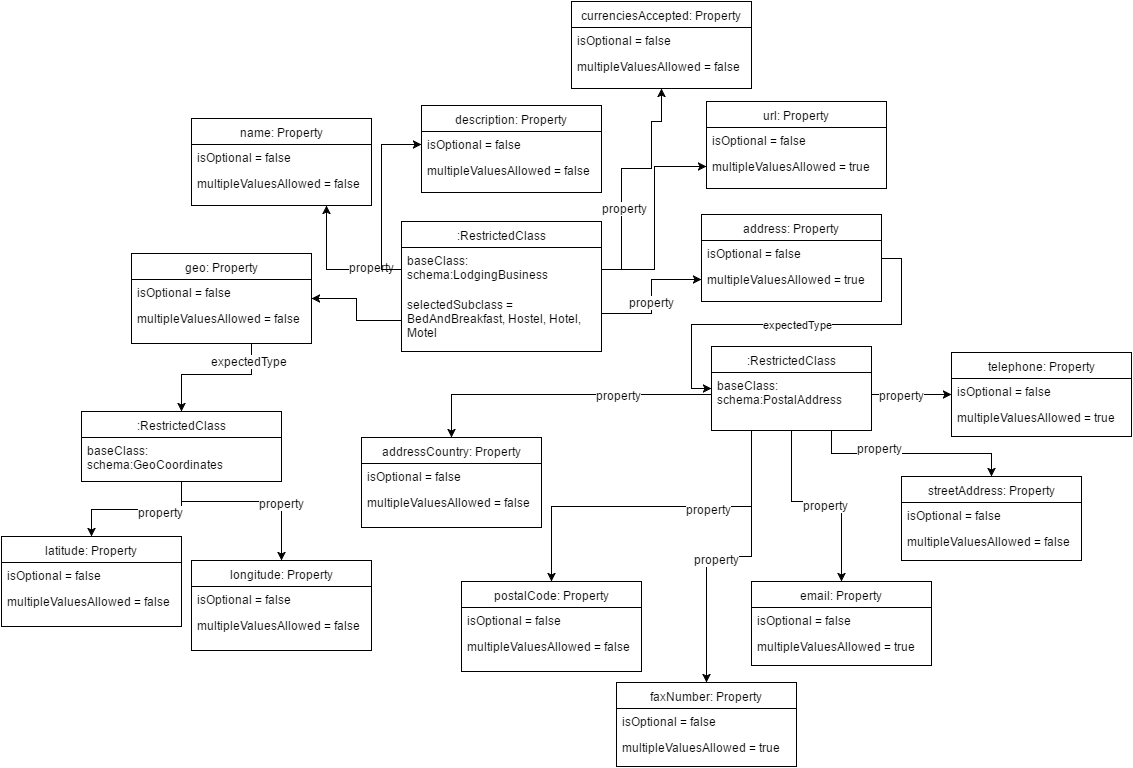}
	\caption{A domain definition for lodging businesses}
	\label{fig:usecase-domain}
\end{figure}

In order to demonstrate our approach and implementation, we created the domain represented in Figure \ref{fig:usecase-domain} and semantic validation rule in Listing \ref{listing-global-rule} via the domain definition interface and rule designer depicted in Figure \ref{fig:ruledef-1}.

In our scenario, a user wants to validate the annotation for Moosleite in Mayrhofen (Listing \ref{lstanno}) against the domain specification and semantic validation rule.
When the user enters the domain specification and rule set to the validator and then validates the annotation, she receives a completeness error. This is because the domain requires the \textit{currenciesAccepted} property but the annotation does not have it.

After the addition of the missing required property to the annotation, the rule-based validation takes place.
The semantic validation rule validates whether the country code of the phone number is consistent with the country of the address. Since this is not the case, the user receives the \textit{"The international country code of the phone number of the place is not consistent with the country of the address."} error message defined in the action part of the rule in Listing \ref{listing-global-rule}.
When the country code of the telephone number is also corrected, the user receives the confirmation that the annotation is valid.

\begin{minipage}[c]{\textwidth}
	\begin{lstlisting}[caption=An example annotation of Moosleite Hotel Mayrhofen. The country code of the phone number does not match the country of the address and the currenciesAccepted property is missing., label=lstanno, captionpos=b]
	{
	"@context": "http://schema.org",
	"@type": "LodgingBusiness",
	"url": [
	"http://www.tiscover.com/moosleite",
	"http://maps.mayrhofen.at/?foreignResource=E33CFC29
	-050E-43D7-9BB3-EA937D33FCA4"
	],
	"address": {
	"@type": "PostalAddress",
	"postalCode": "6290",
	"streetAddress": "Neu-Burgstall 318",
	"addressCountry": "AT",
	"telephone": "+42 5285 62894",
	"email": "eberl.friedl@tirol.com",
	"faxNumber": "0043 5285 62064",
	"url": "http://www.tiscover.com/moosleite"
	},
	"name": "Moosleite",
	"description":  "Our house is situated approx. 1.5km from Mayrhofen, at the edge of the forest and enjoying wonderful panoramic views.",
	"geo": {
	"@type": "GeoCoordinates",
	"latitude": "47.1862746335978",
	"longitude": "11.8581855297089"
	}
	}
	\end{lstlisting}
\end{minipage}

\section{Conclusion and Future Work}
\label{sec:conclusion}
The web we know is changing and the only way to remain visible on the new layer of the web is providing semantically described structured data. Schema.org is helping us to achieve this goal since 2011 as the de facto standard for describing things on the web.

We acknowledge that schema.org adopts "some data better than no data" motto and its data model is imperfect by its nature\footnote{http://schema.org/docs/datamodel.html}. However, it is still important to publish high quality structured data that conforms to the schema.org vocabulary. We aim to help users for achieving this goal with our domain specific validation approach.
In this paper, we introduced a domain specific approach to validate schema.org annotations. Our approach allows domain experts to specify a domain based on a subset of schema.org vocabulary as well as validation rules for semantic consistency. We showed the web based implementation of our approach alongside a use case in the tourism area.

For the future work we will follow the works of different groups, especially the RDF  Data Shapes Working Group, to find out possible alignments between our approaches. For instance, development in the SHACL shows promising results and can be utilized for the later implementation of our approach.

Moreover, we are in the processes of advancing the tool that implements our approach while including the development of more sophisticated rule designer and validator. We will test our tool in a larger scale in tourism domain within the next months.

Our approach currently does not consider multi-typed entities, which are encouraged by the schema.org initiative. For instance, the schema.org hotel extension \cite{karle2017extending} suggests that a lodging business should define their rooms as both \textit{schema.org/Room} and \textit{schema.org/Product} in order to conform schema.org specifications. We will investigate how we can adopt the multi-typed entity notion in the future work.

%
%
\bibliographystyle{splncs03}
\bibliography{references}

\end{document}